\documentclass[manuscript]{acmart}

\usepackage[capitalise]{cleveref}

\AtBeginEnvironment{quote}{\par}
\setcopyright{none}

\begin{document}

\title[AAC in Social Communication and Community Engagement: Experiences and Opinions of Autistic Adults]{The Role of AAC in Social Communication and Community Engagement: Experiences and Opinions of Autistic Adults}

\author{Blade Frisch}
\email{bwfrisch@mtu.edu}
\affiliation{
  \institution{Michigan Technological University}
  \city{Houghton}
  \state{Michigan}
  \country{USA}
}

\author{Betts Peters}
\email{petersbe@ohsu.edu}
\affiliation{
  \institution{Oregon Health \& Science University}
  \city{Portland}
  \state{Oregon}
  \country{USA}
}

\author{Keith Vertanen}
\email{vertanen@mtu.edu}
\affiliation{
  \institution{Michigan Technological University}
  \city{Houghton}
  \state{Michigan}
  \country{USA}
}

\begin{abstract}
    Little research has explored the communication needs of autistic adults. Augmentative and alternative communication (AAC) can support these communication needs, but more guidance is needed on how to design AAC systems to support this population. We conducted an online, asynchronous, text-based focus group with five autistic adults to explore their social communication and community engagement and how AAC might support them. Our analysis found 1) participants' emotional experiences impact the communication methods they use, 2) speaking autistic adults can benefit from AAC use, and 3) autistic shutdown creates dynamic communication needs. We present implications for AAC interface design: supporting communication during shutdown, indicating communication ability, and addressing the fear of using AAC. We provide themes for future autism research: exploring the impact of a late diagnosis, understanding communication needs during shutdown, and researching the social and environmental factors that impact communication. Finally, we provide guidance for future online focus groups.
\end{abstract}



\maketitle

\section{Introduction}

The Centers for Disease Control and Prevention estimates that 1 in 36 children has an autism spectrum disorder (ASD) diagnosis \cite{maenner_prevalence_2023}. Autism is characterized, in part, by a disability in social communication and interaction \cite{american_psychiatric_association_diagnostic_2022}. This communication disability can manifest differently across autistic people.  Some use speech as their primary means of communication, while others speak rarely or not at all (i.e., nonspeaking). Many autistic people use augmentative and alternative communication (AAC) to help meet their communication needs \cite{ganz_aac_2015}, especially those who are nonspeaking \cite{tager-flusberg_minimally_2013}. However, little AAC and autism research has included adults. \citet{holyfield_systematic_2017} found that only 6 of the 18 studies they identified had adult participants and none focused on social communication and community engagement.

Autistic adults, including those who use speech, can benefit from using AAC. However, they may face barriers that prevent them from using AAC, such as a lack of knowledge about the existence of AAC or believing that it is only for nonspeaking people~\cite{zisk_augmentative_2019}. There is a need to conduct more research with autistic adults on their communication needs and how AAC can support those needs, as shown by~\citet{holyfield_systematic_2017} and called for by~\citet{zisk_augmentative_2019}. Additionally, there is a need to research supporting communication beyond what is required to meet basic wants and needs, as called for by \citet{holyfield_systematic_2017} and \citet{light_changing_2012, light_communicative_2014}.

We conducted a focus group with autistic adults with the goal of answering the following research questions:

\begin{enumerate}
    \item How do autistic adults define and engage with their community?
    \item How do autistic adults' current communication strategies impact their engagement?
    \item What do autistic adults want to see in future communication supports?
    \item How can focus groups be made accessible to AAC users and autistic participants?
\end{enumerate}

In our online, asynchronous focus group, we captured the lived experiences of autistic adults as they engaged with their communities using various communication methods, including speech, low-tech AAC, and high-tech AAC. We asked participants questions about their current communication strategies, how these strategies helped or hindered their community engagement, how their communication style changed based on the surrounding context, and what they wished others knew about communicating with autistic adults. 

Our analysis identified themes stemming from participants' internal emotional state, past interpersonal experiences, and external factors such as the actions of communication partners. While many of our participants used speech to communicate, there was a desire to introduce part-time AAC use into their daily lives. However, there was also a fear of how they would be perceived using both speech and AAC. We found autistic adults can have dynamic communication needs stemming from the shared experience of autistic shutdown. Participants suggested that AAC systems may need to provide more focused support during these times of shutdown to help communicate their current abilities to communication partners. Finally, we suggest themes for future AAC design research and provide guidance on how to make focus groups more accessible for autistic adults and AAC users.

\section{Related Work}

\subsection{Autism and Social Communication}

ASD is characterized, in part, by deficits in social communication~\cite{american_psychiatric_association_diagnostic_2022}. As such, there has been much research on teaching social norms and communication to autistic children and young adults~\cite{gates_efficacy_2017, laugeson_abcs_2014, laugeson_randomized_2015, watkins_review_2015}. As discussed by~\citet{zisk_augmentative_2019}, little research has been done on the social communication needs of autistic adults. However, autistic adults have higher levels of loneliness compared to neurotypical adults~\cite{ee_loneliness_2019, petty_blog-based_2023}, which may be related to challenges with social interaction.

One technique autistic people use to try and fit in with neurotypical people is \textit{masking}. When autistic individuals mask, they try to suppress the neurodivergent aspects of themselves to appear more like neurotypical people~\cite{miller_masking_2021}. They often mask to combat the stigma that comes with being disabled~\cite{alaghband-rad_camouflage_2023, hull_putting_2017}. Controlling one's presentation of self can help disabled people develop a sense of community where it might otherwise be lacking. Being disabled can drastically impact one's ability to engage with others~\cite{goffman_stigma_1963}, leading to increased feelings of loneliness, anger, stress, and anxiety~\cite{cacioppo_loneliness_2006}. However, autistic people still desire to be known and understood as they are, not as their masked selves~\cite{frost_i_2019}.

Humans are communal creatures and desire to be in community with each other, including autistic people~\cite{gillespie-smith_spectrum_2024}. Community has been defined by~\citet{cohen_introduction_1990} as ``a specific group of people who reside in a distinct geographical area, share a common culture, and are capable of acting collectively towards a given goal,'' and social communication is essential to this sharing and collaboration. In the age of social media, a community can transcend geographical boundaries~\cite{gruzd_enabling_2013}. Online communities form around a common culture and are capable of acting collectively towards a common goal, but function differently from communities based on in-person interaction and place different communication demands on community members, which may support social communication for autistic people. Online communities often involve asynchronous, text-based communication and have been used by autistic people for finding community~\cite{gillespie-smith_spectrum_2024}. 

\subsection{Augmentative and Alternative Communication}

AAC addresses the needs of people with speech and communication disabilities using a range of tools, techniques, and supports \cite{beukelman_augmentative_2020}. AAC can be broadly organized into two categories: unaided and aided. Unaided AAC is where someone makes use of only their body and does not use external tools. This includes techniques such as signing, gestures, facial expressions, and other forms of movement. Aided AAC makes use of external tools. These tools can be low-tech tools, such as pen and paper, or high-tech tools such as an electronic speech-generating device.

The advent of mobile technology has helped high-tech AAC grow in availability and power~\cite{mcnaughton_ipad_2013}. However, communication support decisions are sometimes based solely on the technology available, with little regard for the needs and preferences of the user. It is important to keep the communication needs of the AAC user as the primary focus during the AAC intervention process~\cite{mcnaughton_building_2019, mcnaughton_ipad_2013, light_putting_2013}. Including AAC users in the design of high-tech AAC systems is also important. \citet{pradana_dialogue_2025} found that excluding AAC users from the design process contributed to AAC's low adoption rate. Pradana also found that AAC, rather than being just a communication mode, can become an integral component of social experiences. 

Different people will have different communication needs that AAC should support, such as social interaction, work activities, and community activities~\citet{light_changing_2012}, as well as different preferences for how their AAC tools should look and function. A variety of AAC interfaces have been created to meet these needs and preferences. Grid displays organize content into rows and columns, where each item typically consists of a symbol or text~\cite{beukelman_augmentative_2020}. Grid displays are used by both children and adults with developmental disabilities~\cite{light_designing_2019, light_new_2019}. Visual scene displays (VSDs) make use of images to support communication~\cite{blackstone_visual_2004}, with embedded hotspots that trigger the system to speak an utterance~\cite{light_designing_2019}. VSDs are especially effective with beginning communicators~\cite{beukelman_augmentative_2020, blackstone_visual_2004}. Other AAC system designs have targeted specific settings, such as dining~\cite{mccoy_implications_2010, obiorah_designing_2021}. Other research has focused on integrating specific technologies, such as context-aware computing~\cite{kane_what_2012}, image recognition~\cite{kane_lets_2017}, and natural language processing~\cite{dempster_automatic_2010, reiter_using_2009, higginbotham_application_2012}.

\subsection{AAC and Autism}

Much research on AAC and autism has focused on AAC use by autistic children. A systematic review and meta-analysis~\cite{ganz_participant_2023} included research with participants up to 21 years of age but reported a median age of 5. Other research involving autistic children has explored the needs of minimally speaking children~\cite{sterrett_effect_2023, chenausky_review_2023, tager-flusberg_minimally_2013}, supporting early social interaction~\cite{charlop-christy_using_2002, laubscher_effect_2019}, and developing speech~\cite{tager-flusberg_minimally_2013, mirenda_autism_2013} and literacy~\cite{caron_effects_2018} skills.

However, recent reviews and meta-analyses~\cite{ganz_aac_2015, ganz_meta-analysis_2012, holyfield_systematic_2017} show that few studies have explored the impact AAC can have on the communication of autistic adults.~\citet{zisk_augmentative_2019} show that part-time AAC use can be beneficial to speaking autistic adults. \citet{hanson_my_2014} showed that AAC can serve as a supplemental strategy for people whose speech is not always intelligible. However, a lack of AAC knowledge and a belief that the ability to use speech means that AAC use is not acceptable can present barriers to AAC use by speaking autistic adults~\cite{zisk_augmentative_2019}.

Recently,~\citet{martin_bridging_2024} conducted semi-structured interviews with autistic adults to explore their use of AAC. They found that typing is faster but symbol-based systems require less mental effort, using AAC can sometimes feel unsafe, and accessing AAC intervention as an adult can be difficult. Other research has explored how autistic people communicate using more general-purpose technology, such as remote work environments~\cite{das_towards_2021} and video calling~\cite{zolyomi_managing_2019}. 

\section{Methods}

\subsection{Research Design}

We chose a focus group~\cite{krueger_focus_2015} as the research method to gather qualitative data on the lived experiences of autistic adults and their experiences with social and community engagement. A focus group allows participants to interact with each other, building up shared ideas and contrasting their lived experiences. Such data on compared life experiences cannot be achieved through methods that focus on individuals, such as interviews and surveys. Autism presents differently in everyone, so each participant will have unique experiences that can also overlap with those of other participants. A focus group allows us to gather both the experiences of the individual participants and how these experiences compare to those of other participants. This study was approved by the first author's Institutional Review Board.

\subsection{Participants}

Participants had to be: (a) on the autism spectrum, (b) 18 years of age or older, (c) fluent in English, (d) able to access the internet, use a web browser, and post messages in a forum. Recruitment was limited to the United States and was conducted through word-of-mouth and advertising to autism advocacy and support organizations.

~\citet{krueger_focus_2015} and~\citet{goodman_observing_2012} both recommend that focus groups have 4--6 participants for participants to feel comfortable and to collect more in-depth responses. Following this advice, we recruited five participants. We do not present per-participant demographics as several participants requested reporting only in aggregate form. Participants ranged in age from 23--38 years old. Four identified as female and one identified as non-binary. The lack of male-identifying participants was not an intentional choice but simply reflects the gender identities of those who expressed interest in our study. Four participants self-disclosed receiving an autism diagnosis as a teenager (\textit{n}=1) or as an adult (\textit{n}=3). The remaining participant elected not to share their age of diagnosis. Participants used the following communication strategies: speech (\textit{n}=4), communication book (i.e., low-tech AAC)  (\textit{n}=1), speech-generating mobile applications (i.e., high-tech AAC)  (\textit{n}=2), and American Sign Language (\textit{n}=1).

\subsection{Focus group platform and structure}\label{platform_selection}

We conducted an online, asynchronous, text-based focus group using the Flarum\footnote{\url{https://flarum.org/}} forum software hosted on FreeFlarum.\footnote{\url{https://freeflarum.com/}} Given that some participants were AAC users, we needed to make sure that the focus group was accessible to them. Previous research has used a text-based medium for focus groups with AAC users~\cite{caron_social_2017, dattilo_i_2008}. This allowed the focus group to be asynchronous, giving the participants as much time as they needed to process and compose messages. As discussed by~\citet{beneteau_who_2020}, asynchronous communication also opens qualitative research up to individuals who may not be otherwise able to participate.

We made our focus group anonymous to protect the identities of our participants. Each participant was assigned an alias of the form P\#. We also wanted to prevent direct contact between participants outside of the forum posts. We included the following Flarum plugins:

\begin{itemize}
    \item Approval
    \begin{itemize}
        \item Prevents participants from creating their own discussions.
    \end{itemize}
    \item Flags
    \begin{itemize}
        \item Allows users to flag posts for admin review.
        \item We instructed participants to flag any posts they thought were inappropriate. This feature was not used, but providing it was part of our strategy to protect participants.
    \end{itemize}
    \item Mentions
    \begin{itemize}
        \item Allows users to tag another user with @Username and cause a notification to that user.
        \item The participants were told about this feature but did not use it, preferring to reply directly to posts.
    \end{itemize}
    \item Subscriptions
    \begin{itemize}
        \item Allows users to follow discussions and receive notifications for new posts and replies in the discussion. \item While this feature was mentioned to the participants, they did not use it. We discuss this further in \cref{discussion}.
    \end{itemize}
    \item Suspend
    \begin{itemize}
        \item Allows admin to suspend users from posting, including posting replies.
        \item This feature was not used, but providing it was part of our strategy to protect the participants.
    \end{itemize}
\end{itemize}

Following guidance from \citet{krueger_focus_2015}, the first author acted as the moderator and created a discussion board for each week of the study. We pinned a post that contained the prompts for the participants to respond to at the start of each week, for example: ``Could you share some stories about how you engage with your community?'' (see \cref{appendix_prompts} for all our prompts). 

Participants were required to respond to the weekly prompts and were also encouraged to respond to other participants' posts. We instructed participants to make their replies within the week but allowed participants to continue their discussions with each other even after the next week opened. Participants were paid \$40 per completed week for a maximum payment of \$200. The weekly themes were:
\begin{enumerate}
    \item Week 1: Becoming familiar with the platform
    \item Week 2: Sharing stories about community engagement and the impact of their communication methods on that engagement
    \item Week 3: Sharing how environmental and interpersonal factors impact their communication
    \item Week 4: Sharing what they would like to see happen in the future
    \item Week 5: Providing feedback on the focus group methodology
\end{enumerate}

We reviewed all new posts on the forum twice a day. Participants were sent a reminder email with a direct link to the week's discussion post three days before, one day before, and on the day a week's discussion replies were due.

\subsection{Analysis}

We used a two-stage analysis process on participants' posts. To prepare for analysis, the first author read through all responses several times in what \citet{tracy_qualitative_2020} called the data immersion phase. Next, the first author developed first-cycle codes using the framework of~\citet{saldana_coding_2021} and applied these codes to phrases, sentences, and paragraphs using a single-coder approach. The first author then used pattern coding \cite{miles_qualitative_2020} on these first-cycle codes to create categories. The pattern coding process was then reapplied to these categories to create themes.

Following the process of \citet{creswell_qualitative_2023}, the second and third authors acted as peer reviewers, asking questions about the analysis and interpretation throughout the process. This is done to ensure rigor in the qualitative analysis process \cite{creswell_qualitative_2023}. Additionally, the first author conducted a member reflection~\cite{tracy_qualitative_2020} with each participant to explain the analysis process and our findings. During this reflection, the first author presented the analysis process, analysis results, and the discussion section of this paper. The participant was then asked to provide feedback on whether they were accurately represented and to provide any other feedback.

\begin{table}[tb]
    \caption{The themes and the categories of responses that make up each theme found during our analysis of focus group posts.}
    \label{table_themes}
    \begin{tabular}{ l l }
        \toprule
        Theme & Pattern categories \\
        \midrule
        Internal life         & Negative Emotions \\
                              & Positive Emotions \vspace{2mm} \\ 
        Interpersonal life    & Barriers \\
                              & Management strategies \\
                              & Relationships \vspace{2mm} \\ 
        External life         & Impact of context \\
                              & Impact of communication partners \vspace{2mm} \\ 
        Future life           & Status indicator \\
                              & ``Say what you mean'' \\
                              & Being human \\
        \bottomrule
    \end{tabular}
\end{table}

\section{Results}

Participants wrote a total of 64 posts, ranging from 9--16 posts per week. The posts consisted of 20,152 words, with an average post length of 286 words and a median of 157. As shown in \cref{table_themes}, four themes emerged from our analysis of these posts: 1) internal life, 2) interpersonal life, 3) external life, and 4) future life. We present these themes and pattern categories in detail below, with representative quotes from participants. We use terminology recommended by \citet{zisk_what_2024} (e.g., ``nonspeaking'') when summarizing our analysis, but preserve the original terminology used by the participants (e.g., ``nonverbal'') when sharing their quotes.

\subsection{Internal Life}

\textbf{Summary:} Participants wrote about the emotions they feel when engaging with others and their community. They shared more negative experiences than positive experiences, but wrote about both. Several participants shared stories of fear, failure, stress, and feeling misunderstood when interacting with others, sometimes leading to shutdown. However, they also wrote about their journey towards self-understanding, how they have learned to unmask and present more authentic versions of themselves, and the impact being understood had on their internal life.

\subsubsection{Negative Emotions}

Several participants shared negative experiences related to social communication and community engagement. Some participants spoke of a \textbf{fear of using communication supports} when interacting with others. P2 described feelings of anxiety surrounding using AAC:
\begin{quote}
    Having friends and people who encourage these methods of communication and understand my communication needs has helped me unmask somewhat in this area, but I still struggle with a lot of compulsive masking and social anxiety that hinder me from adapting to my own needs more regularly.
\end{quote}
P5 described their fear as a lack of bravery: ``I have a communication book that I carry with me, but I've not been brave enough to use it.''

They also stated their fear includes a \textbf{fear of being misunderstood} and of identifying as disabled: ``Sometimes I want to really really own my disabilities with autism. Not just carry a little book in my pocket that I'm terrified to use because I don't think anyone will understand.'' P5 further explained this can extend to \textbf{feelings of failed communication}. They framed this feeling in the context of a former work manager who did not understand communicating with an autistic person, sharing ``I had tried and struggled constantly with the previous manager to communicate with him ... It's just left me feeling so [bad] that I don't know how to communicate.''

These feelings contributed to \textbf{feelings of high stress}. P3 outlined experiences with their family:
\begin{quote}
    I've often found myself overwhelmed and overstimulated, unable to do anything about it because I'm trapped in a situation with them, and any attempt to remove myself or accommodate my needs would be met with resistance and outrage, and a complete lack of empathy.
\end{quote}
A \textbf{lack of understanding from others} also contributed to these heightened stress levels. P2 shared in another response, ``My family knows I identify as autistic but don't all fully understand or accept it, which leads to a weird combination of unmasking and masking.''

When participants were highly stressed, it often led them to \textbf{shutdown}. P3 stated, ``Until fairly recently I didn't actually realize or understand that I go nonverbal fairly often, that I have shutdowns, and that I have real difficulty when I have to speak in real-time, out loud, with another person.'' P2 shared similar struggles with their family:
\begin{quote}
    My biggest struggle with communicating is that when I start to have a struggle or feel I'm not being understood or heard, I'm more likely to shutdown than to actually try to communicate, because the difficulty of the struggle is just not worth it a lot of the time.
\end{quote}

\subsubsection{Positive Emotions}
Participants also shared positive experiences with social communication and community engagement. P4 told a story about how working at an organization for autistic adults helped them gain \textbf{self-understanding}:
\begin{quote}
    I learned a lot more about myself and my disability and was no longer ashamed. I saw individuals that had more severe autism than myself and was grateful and proud that I was able to achieve the things I had achieved thus far. Plus, the staff that worked there were trained in autism and were able to help me understand my diagnosis and help me become a better me.
\end{quote}
P5 also shared their journey of self-understanding: ``Knowing I was autistic in the last year, it's a pretty recent diagnosis, helps me understand that it's important for me to not just advocate for myself but use tools that are available to me.''

Participants also described the importance of \textbf{being understood by others}. P3 shared how important understanding friends can be: ``Most of my closest friendships over the years have been with people who share some of my struggles with communication, or at least understand and empathize with me and do not punish me or make me feel badly about my struggles.'' P2 echoed this sentiment, sharing ``These two [close friends] are the only people in person who know I'm autistic, believe it, and actually accommodate me.'' P4 extended this to their family members being understanding and accepting: ``When I was first diagnosed, I found it easier to be myself around my family and friends and unmask.''

\subsection{Interpersonal Life}

\textbf{Summary:} Participants faced barriers related to social norms, processing time, and the effort that comes with communication. They also wrote about developing a series of strategies to navigate these barriers, and reflected on how these barriers and management strategies had both positive and negative influences on their interpersonal relationships.

\subsubsection{Barriers}
P4 shared stories of how \textbf{not understanding social norms} impacted their interactions, stating:
\begin{quote}
    I tend to have one-sided conversations, where I do most of the talking. It took me a long time to realize that I needed to let the other person talk some as well ... Sometimes I talk a little too much, and people tend to avoid me in the future when they see me. It makes me sad because I wish I had just been told that I was talking too much so I could work on it.
\end{quote}
P4 also shared how their verbal interactions at work led them to feelings of isolation:
\begin{quote}
    I work at a grocery store and I communicate with my co-workers and the customers verbally on an almost daily basis. I need to verbally communicate with customers in order to bag the groceries the way they would like ... Sometimes I talk a little too much, and people tend to avoid me in the future when they see me. It makes me sad because I wish I had just been told that I was talking too much so I could work on it.
\end{quote}

P1 wrote about the impact of \textbf{processing time} on their communication: ``[if] everyone is moving at a slower pace [this] makes a big difference in order for me to find my communication at its best.'' P1 shared this also happens when interacting with communities on Discord: ``I find it both really important to be in these discord servers and really difficult and usually end up lurking and not talking much because it takes me so long to process that by the time I do things have moved on.'' P2 shared this sentiment: ``I do think it would be helpful for them to be aware that sometimes it just takes longer to process things and a lack of answer doesn't mean we're ignoring them or don't understand.''

Another common barrier described by participants was the \textbf{effort required for communication}. P1 shared:
\begin{quote}
    I'm going to be exhausted and find communication hard. Things making it easier doesn't change that and it being hard doesn't make my communication methods hinder me. I do often find so much hard not because [my communication method] is hindering me, but because even with all the benefits and advantages and improvements it is just plain still hard.
\end{quote}
P2 replied: ``Even though I can communicate verbally, and effectively too, it's extremely overstimulating for me and I burn out so quickly and struggle to maintain relationships where I'm always masking or using communication methods that don't suit me.''

\subsubsection{Management Strategies}

In order to minimize the barriers they face, participants described various management strategies. Some participants opted to modify their communication strategies for the \textbf{comfort of the communication partner}. P5 explained this as a fear response, stating ``In general, I haven't been very brave in utilizing other communication resources that I have. I tend to rely on what is comfortable for the other individual.'' P2 shared a similar fear: ``Being so high-masking I also have to unwork a lot of my people-pleasing tendencies, and this hinders me from using my preferred methods of communication just because I think someone might perceive it negatively, without even trying first or necessarily having evidence to support this.''

Some participants also shared moments of \textbf{self-advocacy}, despite the fear, to manage their communicative ability. P3 described a routine they implemented at home to manage their stress:
\begin{quote}
    I've started asking for two hours of nonverbal time to myself each day after I get up, take care of my mom and our two dogs, and go through my morning routine. I can retreat to my bedroom and have control over what I'm doing and the level of sensory input I'm receiving ... The critical bit is that I control who has access to me during this time, and how much communication I have to process. If I'm feeling overwhelmed and processing speech is difficult, I can limit the amount of words coming into my ears and listen to some instrumental music.
\end{quote}

One strategy used by the participants was \textbf{conversation scripts}. P3 described their method as ``If it's a phone call with someone I don't know, a place I'm unfamiliar with, or for something important, I have to script out the conversation as much as possible before I pick up the phone to help ease the anticipatory anxiety.'' P1 wrote how they anticipate conversations and prepare messages in their AAC system:
\begin{quote}
    I am able to use some degree of saved phrases in my AAC in order to help with processing ahead of time - this is incredibly helpful for doctors for example; plan everything I am going to say, and save as many different possible pathways conversations could go - but this is of limited use just because of the way that communication is itself an interaction between multiple people.
\end{quote}

Another common strategy for managing communication was making use of \textbf{asynchronous communication} methods. P1 described the benefits of asynchronous communication: 
\begin{quote}
    I find expressive and receptive communication far easier in async settings. [It makes a big difference] having time to process as well as the time to be able to communicate in a manner which works best for me rather than how I am best able to in the moment.
\end{quote}
P2 agreed, stating ``Asynchronous text-based communication has always been easiest for me too.'' P2 wrote in another post:
\begin{quote}
    When I am overstimulated I prefer written methods of communication, though usually in social settings my masking will override my needs and lead me to burning out later. Because of this, many of my closest friendships have been online or with people who I largely communicated with over text/written messages. This allows me time to process and think of the right thing to say, and I don't have to focus on facial expressions, body language, etc.
\end{quote}
These quotes show that participants valued the ability to control their expression of self with others, and having the time to process the communication of others and prepare their own communications affords them that ability.

\subsubsection{Relationships}

While these management strategies helped participants make and maintain relationships, they also shared \textbf{feelings of not belonging}. P1 described it as ``I've found really finding and belonging in community is complicated, and I don't feel like I truly belong anywhere. Part of this is because of communication being difficult.'' P3 described their experiences with online communities as ``I just feel like an outsider, like I'm intruding. And I feel awkward if I try to insert myself. So many attempts at participating in conversation are often halting and unsure, if I even try at all.'' P4 told of their experiences working in a grocery store and what it is like interacting with customers, succinctly stating ``I just want to feel liked and fit in, just like everyone else.''

Participants described \textbf{wanting connections with others, to make friends, and to be accepted}. P3 shared:
\begin{quote}
    One thing I do know, however, is that I often experience this feeling of wanting desperately to reach out and connect with the people in my life and build stronger bonds with them, but am often unsure how to do so. Some piece of the puzzle of making friends just seems to elude me. Or, a group of people in one of my communities will be getting together to hang out, but it will be in a voice chat or in person and I know that that's beyond my capabilities in that moment, and so I don't participate. That's always really lonely. I want to, but I don't know how to engage in these kinds of situations.
\end{quote}
P4 confirmed they have similar feelings, replying ``I am the same way and it's difficult to start and maintain friendships.'' Participants want to engage with others, take part in social communication, and engage with their communities, but it can be difficult for them to do so. P1 stated, ``I also have people I'm only in contact with via social media and would like to have other contact with but it is just too difficult to navigate.'' P2 shared ``I repeatedly find myself with close friends who I only know online or who are long-distance because of this, even though I really crave in-person connection with people.'' This shows the need to support in-person social communication as well as digital communication.

\subsection{External Life}

\textbf{Summary:} The participants wrote about aspects of social communication and community engagement that they had no direct control over. They shared the impact of the physical environment, and of other people's perceptions and attitudes towards autism. These perceptions sometimes led to others taking harmful actions toward the participant. However, the benevolent actions of other people also could support participants in being more authentically themselves.

\subsubsection{Impact of Context}

Participants talked about how the \textbf{environment} can impact their communicative capacity and abilities. For example, P1 explained ``Different environments require different things. If it's raining, I'm not going to be able to use something with a capacitive screen. I can't use eyegaze when I'm walking around. Not everything is always going to be accessible, not to mention easiest.'' P3 wrote about their experience attending a convention and the impact the environment had on their communication:
\begin{quote}
    At my first anime convention, the busiest day was really hard for me ... I had other bodies pressing in on all sides, and could feel other peoples' sweat on my skin, and it was a LOT for me to handle. I felt like a very small child in those moments, and was actually unable to communicate with words. It was purely me tugging on the part of [my partner's] shirt I was holding onto, or clenching his hand tighter.
\end{quote}

The social environment can also affect communication. P2 wrote about the impact of the \textbf{number of communication partners}:
\begin{quote}
    In terms of the number of people I'm communicating with, I am typically much more talkative and better at communicating with just one or two people, three on a good day, and once there are more people in a conversation I typically just listen and do not contribute to the conversation, as it is overstimulating to process.
\end{quote}
P3 shared a similar sentiment:
\begin{quote}
    When I'm talking to someone one-on-one I have a much easier time. The more people you introduce into the conversation, the harder it gets to keep track of not only what they're saying, but what they mean by what they're saying. If it's in person, then you add a whole bunch of other physical signals too, and have to sort through tone and volume and other things that you don't have to think about if you're talking in writing.
\end{quote}
As the number of communication partners increased, so did the difficulty of actively engaging in conversation. For both P2 and P3, there was a point where they became too overwhelmed to effectively communicate and they transitioned to a more observational role.

The environmental and social factors combine when \textbf{communities make themselves accessible} to others. Participants shared groups they identified as their communities, with many being digital communities. They also shared how accessible these communities were to them. For example, P3 shared:
\begin{quote}
    I often hang out in a voice chat in Discord with some of my friends, but this can be very overwhelming and sometimes I have to find a way to leave because I can feel myself starting to lose the ability to think clearly and speak those thoughts out loud.
\end{quote}
Online communities can be more accessible when there are options for interaction to choose from based on the participant's communication needs. For example, P2 wrote:
\begin{quote}
    I am much more free in these [Discord servers] to communicate how I want and on my own time, and can choose whether to type or send audio messages as suits my communication needs at the time. 
\end{quote}
Even if the number of communication partners and the environment are favorable for the participants, their stories show it also takes intentional effort to make a community accessible to autistic adults.

\subsubsection{Impact of Communication Partners}

Some participants expressed that it was \textbf{easier to communicate with familiar people}. P1 described this as ``If I am just talking to close friends then I don't need to put nearly as much effort into carefully editing every word for hours in scripts before meetings.'' If there were issues with the closer people accepting the participant, however, then the communication was more difficult. P3 described their experience with unaccepting family after being diagnosed as an adult as ``I think the fact that people I already know had certain expectations; they know a certain version of me. And that version has been heavily masking my entire life.''

Other participants found it \textbf{easier to communicate with strangers}. With people they were unlikely to see again, they found it easier to identify as disabled, ask for accommodations, and communicate in ways that were more comfortable for themselves and their communication partners. P1 shared: ``So often people act like how well you know someone is related to how communication works in ways that assume strangers are inherently less comfortable, yet the `I'll never see this person again' is itself very freeing.'' P2 reported seeking out strangers, stating: 
\begin{quote}
    I really like going and doing stuff out in public by myself for this reason, and I actually like meeting new people to some degree because often I'm less masked and more comfortable talking to them for the first time, and after that is when I start to feel awkward and hyper-analyze my conversations.
\end{quote}
P3 used this as a way to become more comfortable identifying as disabled in what they considered low-stakes circumstances, sharing ``I've gotten more used to and more comfortable with applying the words [I'm disabled] to myself and asking for some basic level of accommodations.''

Another factor is communication partners' \textbf{perceptions of and attitudes about autism and autistic adults}. P1 wrote how this impacts their safety, sharing ``It is not safe for me to communicate in a `more disabled' looking manner to people who have the power to do things that could cost me too dramatically much.'' P3 shared their experience with how their family responded to their autism diagnosis: 
\begin{quote}
    My family had much the same reaction [to my diagnosis], except it didn't come from a place of curiosity --- at least one of them flat out told me (and in the case of one family member continue to invalidate my experiences to this day by telling me still) that they don't believe I have autism. It sucks.
\end{quote}

Communication partners' negative attitudes sometimes led to negative interactions. P1 wrote:
\begin{quote}
    [What] hinders me isn't about my communication methods. It's about how other people treat me. Other people not listening to me isn't there being anything wrong with my communication methods. Other people talking over me isn't there anything wrong with my communication methods. Other people moving too fast for me isn't having anything wrong with my communication methods.
\end{quote}
This indicates that no amount of communication support can make up for a lack of understanding and accommodation from others. While providing communication support like AAC can help address many barriers to communication, autistic adults can still face barriers simply for existing as they are.

\subsection{Future Life}

\textbf{Summary:} Finally, participants were asked to describe what they would like to see happen or be created to better support their social communication and community engagement. Some participants responded with a desire to show others their ability to communicate and interact with others. Other participants wrote about how they wished others would modify how they communicate with an autistic person. Nearly all participants shared a desire to simply be viewed as human, with unique communication needs and methods.

\subsubsection{Status Indicator}

Some participants wrote about \textbf{wanting a way to indicate to others} that they are disabled and need communication accommodations, or what communication method they currently prefer. P5 described their solution as:
\begin{quote}
    Instead of just a little book [in] my pocket, maybe something that would be seen that would say hey you need some extra support and help and understanding. You're autistic? Cool. I don't know what that means for you. Can you help me understand so I can help you?
\end{quote}
P3 wrote about working on a status indicator with their therapist:
\begin{quote}
    I have recently come up with a few ideas for physical visual cues to communicate my mental state and my current openness to outside communication to other people without having to use words ... the core idea is some simple color-coded system (think traffic lights, or the little ``vacant/in use'' sign on some bathroom doors) to use on my bedroom door so that the people I live with know if I'm open to them knocking and coming to talk to me or not.
\end{quote}
P3 elaborated in another post:
\begin{quote}
    But for communication in general...I think the support systems I've been thinking about that would help me involve a lot of just...ways to signal to people where I'm at and what I'm feeling when words fail me. Being able to have a standard system in place that the people around me understand so that I can advocate for myself when the ``normal''/acceptable routes of communication aren't an option for me, or when my logical mind takes a back seat and my emotions take over my body.
\end{quote}

\subsubsection{``Say What You Mean''}

Some participants wrote about their experiences with \textbf{others saying one thing and meaning another}. This impacted P5 in their work environment, with a misunderstanding leading to an undertreated injury. Their manager told them to fill out paperwork after they received a cut at work and to take care of the cut. P5 filled out the paperwork but received no direction on how to manage the cut so they bandaged it themselves. They were later scolded by the manager for not going to urgent care, the implied strategy for managing the cut. P5 described the situation as ``I told him I never heard him say any of it. He told me this and so I did this.'' The manager then admitted he should have been more direct in the protocol and following up on the incident.

P3 summarized their experiences in another post: ``I wish people would learn to just say what they mean, and mean what they say. It's so hard to run in circles to decipher the code I was never given a key for.'' These stories show that using implications and indirect communication can lead to personal and even safety concerns.

\subsubsection{Being Human}

The topic the participants labeled as most important to them was \textbf{wanting to be treated like human beings}. P1 described how using AAC to communicate can be perceived by others:
\begin{quote}
    The people I know best, the strangers who will never see me again, and the people who routinely see me but in matters where they are able to make decisions about my life, are all dramatically different relationships which require different sorts of communication. This is even more drastic when I am seen as not human because of how I communicate.
\end{quote}
Just as every person is unique, so is the way autism manifests in each person. P2 shared ``The main thing I would want my community/people in general to know about communicating with autistic adults is that we all communicate differently and it's best to ask the individual.'' P1 expanded on this idea in their response, writing:
\begin{quote}
    Because you really can't generalize between communication with the very extroverted autistic family members of mine and me. You can't generalize between autistic friends of mine who absolutely hate text-based communication and me. I just want to be treated as a person.
\end{quote}

Concerning this, P5 wrote ``We aren't just choosing to be autistic because we feel like it, that we are jumping on some sort of trend. I really truly wish that people would understand that while everyone may struggle with similar things, it is not the same!'' When replying to P1, they stated ``I feel like the world doesn't let us be humans or recognize that we still are.''

\subsection{Focus Group Feedback}

All participants reported that the focus group was a positive experience. P3 shared how validating it was to hear from others: ``Reading other peoples' answers to the prompts was interesting, and led me to further exploring my own struggles in ways that I hadn't thought of before.'' P4 echoes this sentiment, writing: ``I realized that a lot of the things I struggled with were a lot of the same things others with Autism struggled with. It was helpful to hear other's experiences with their Autism and specifically how it related to their communication.'' The ability for participants to interact with each other and receive this validation is an aspect of focus groups that sets them apart from more individual-focused methods such as interviews. It also allows for ideas to build off each other, which other asynchronous methods like surveys do not support.

We observed that participants sometimes struggled to interact with each other. A contributing factor was not being aware of all the features available in the forum software. While some instruction on the forum was provided, more was needed. For example, P2 suggested automated email reminders to supplement the manual reminders sent out by the research team. As discussed in \cref{platform_selection}, we included the Subscriptions plugin to provide this exact feature but participants did not make use of it. P5 shared that they occasionally struggled to know which replies were meant for them, which could be solved by the Mentions plugin.

Almost all participants directly mentioned wanting to keep the online, asynchronous, anonymous format. P1 wrote: ``I would really like more to be done with anonymous focus groups. So many of the focus group I've seen are not anonymous.'' P5 explained that ``[The anonymity] allows me to open up more and give honest replies.'' P2 shared that ``It stimulated good responses that private interviews on the one hand or fast-paced communication like video/audio calls or chats on the other hand would not have.'' P1 wrote about the importance of time gained by being asynchronous: ``Being able to have a text-based format which allows for any amount of thinking between answers is really helpful and allows for so much more in-depth of answers.''

\section{Discussion}\label{discussion}

Emotion played a large role in the participants' communication. Several wrote about managing their communication based on the comfort of the other person, fear of using AAC, and feelings of failed communication. However, participants also shared that their feelings towards their communication, and their disability overall, changed as they worked towards self-acceptance. The impact of the participants' emotional lives affected how much they were willing to use modes of communication other than speech.

Our results show that speaking autistic adults have both a need and a desire to use non-speech modes of communication. This can be especially useful when they begin shutting down. Participants described shutting down when they were overwhelmed or overstimulated, whether from the environment, the actions of others, or when the interaction became too much to process (e.g., their were too many people in a conversation or people were speaking too fast).

When in shutdown, they told of an inability to use speech to communicate. This created a dynamic, temporary communication need; participants would typically emerge from shutdown once the trigger was removed. The dynamic nature of this state means that the communication needs of the participants were also dynamic and could change quickly. This shows that AAC can be a valuable supplemental communication support even for speaking autistic adults.

Some of the management strategies participants shared have been explored previously. For example, some participants wrote about using scripts to prepare for important or potentially stressful interactions. Scripts have been used in AAC research to build up the confidence to use AAC in public~\cite{lasker_promoting_2001}, to teach AAC use through computer-based training for ordering in a dining setting~\cite{mechling_computer-based_2006}, and to support social interaction with coworkers~\cite{heller_promoting_1996}. In this study, we saw both speaking and nonspeaking participants using scripts as a communication management strategy.

We also saw that participants often turned to online communities and asynchronous communication. They described these communities as being an accessible way to maintain relationships with others that can better match their communication needs than in-person interactions. A similar trend has been seen in people with ALS~\cite{kane_at_2017} and adolescents with cerebral palsy~\cite{caron_social_2017}. In these studies, online communities and asynchronous communication were more accessible to AAC users, especially those who used an alternative computer access method such as eye tracking. In our study, participants reported that asynchronous interactions were a better fit for their communication processing needs. Text-based online communities provided opportunities for participants to process interactions on their own time. However, when online communities used speech as the mode of communication, participants shared that they struggled to process the speech with enough speed to keep up with the conversation.

Finally, all participants spoke about the desire to be seen as human. Every autistic person is a unique and individual person. While there are common characteristics inherent to autism, each person will manifest and express themselves differently. It is important to recognize that there is no one-size-fits-all approach to communicating with or supporting the communication of autistic adults and that they should get to know the individual. There is also a need to better educate the general public on what autism is and how it impacts communication.

\subsection{Implications for AAC Design Research}\label{aac_design}

There is a need to design AAC options that support communication when autistic adults are in shutdown. When participants shared their experiences with shutdown, participants described being able to feel themselves going into shutdown and how their communication abilities changed or even disappeared when in this state. They also wrote about the effort that is required to communicate, which is especially true during times of shutdown. However, the communication effort required is dynamic and can change based on the person's abilities in the moment.

AAC systems can be designed to allow users to preplan content on the days they have more energy to use during times of shutdown. This aligns with the conversation scripting management strategy that some participants employed. Such a system could help people plan out critical conversations that could arise while they are in shutdown, scripting out what they need to say as well as replies to anticipated questions when they have the energy for it. As one moved into shutdown, they could read from the preplanned conversation scripts to alleviate the effort required for communication. The system could also provide text-to-speech so the person could still communicate when unable to speak.

Artificial intelligence (AI) could also help with preparing for social communication and community engagement. Generative AI could be used to help prepare an autistic person for planned upcoming conversations. One aspect could be to help generate content for the conversation scripting, reducing the effort of creating the scripts. Another aspect could be using an AI conversational agent to help practice the conversation scripts. Such practice could help a person develop familiarity and comfort with communicating in a new social or community setting. It could also help further refine the conversation scripts, highlighting conversation paths they might not have prepared for or times where they are not fully communicating their desired intent. Additional research is needed to explore the needs and preferences of autistic people with regard to the use of AI to support communication, as well as the ethical and privacy implications.

Our focus group suggests a need to explore a status indicator for communication. There have been examples of indicators with people with ALS \cite{sobel_exploring_2017}, but these indicators were more focused on indicating aspects of the conversation (e.g., conversational turn-taking). Our participants desired a system to indicate their level of communication ability to others so that their communication partners would be more open to modes of interaction other than speech. Research is needed to explore what such a system would look like and how it can be integrated with AAC systems and study the social perception of using a status indicator. Research by \citet{sobel_exploring_2017} showed that people with ALS did not want to be marked as ``other'' by their AAC use. Our participants did write about owning their disability and desiring to be seen as autistic, but the social perception of such a system needs further exploration.

\subsection{Implications for Autism Research}

There is little research that focuses specifically on autistic adults, as much of the current literature focuses on children. As such, there is a general need to conduct research with autistic adults to better understand their unique lived experiences.

Four of our five participants disclosed receiving an autism diagnosis later in life. This created a unique environment where both the participants and the people who knew them before their diagnosis had to come to terms with the diagnosis. We identified examples of self-understanding and how participants learned to begin unmasking, but that was not the focus of this work as no follow-up questions were asked concerning this. The impact of a late diagnosis would be an interesting avenue for future research.

As discussed previously, participants often wrote about the impact of what they self-described as shutdown. Going into shutdown created unique communication needs for the participants, usually different from those they have outside of shutdown. There has been research conducted on the physical nature of shutdown \cite{phung_what_2021, vaquerizo-serrano_catatonia_2022}, but more research is needed to better understand the impact of shutdown on communication specifically.

Finally, participants wrote about the impact of other people on their social communication and community engagement. They shared stories of being misunderstood and of others not believing or accepting their autism diagnosis. The perceptions, attitudes, and actions of neurotypical people towards autism and autistic people need to be better understood. Antagonistic beliefs towards disabled people can be a safety concern, as shared by one participant, and exploring how to reduce the stigma around disability is needed.

\subsection{Implications for Future Online Focus Groups}

Participants reported positive experiences with the online focus group. They explicitly noted that being able to interact with each other would likely lead to researchers collecting richer data. Having the focus group online, asynchronous, and text-based helped to make it more accessible for all participants, including those who used AAC. It gave them time to process responses, formulate their thoughts, and gather the energy necessary to engage with the focus group.

It is critical to provide robust training on how to use the platform when running an online focus group. We instructed participants on the basics of making posts, replying to posts, and tagging other participants. However, we think participants would have benefited from more thorough instruction on features like subscribing to posts and setting up email notifications. We recommended developing more guided exploration and practice activities at the beginning of a focus group to enhance interaction during the remainder of the focus group.

Finally, finding the correct platform for the focus group was an important step in the study design. Flarum proved to be a useful tool due to the ability to customize the platform with our desired features, security and privacy controls, and being free to use. Some of these features, in particular those related to privacy and protection of participants, were crucial for obtaining our ethics approval. The participants also spoke highly of this platform, despite their suggestions for improvement. This software was simple to use and can be easily deployed for future focus group research.

\section{Conclusion}

This study explored the experiences of autistic adults with social communication and community interaction through an online, asynchronous, text-based focus group. We found that participants experienced both positive and negative emotions when engaging with their communication partners. They encountered barriers that inhibited social communication and community engagement but developed management strategies to navigate these barriers. However, their relationships were still impacted by their communication disability no matter how well-managed. This partially stems from the impact that context and other people have on social and community interaction. The participants shared what they would like to see happen or created to better support their social and community engagement.

Our analysis revealed that autistic shutdown has a large impact on the social communication and community engagement of autistic adults. We also found that receiving an autism diagnosis later in life can create a complex sense of self, with late-diagnosed adults often undergoing a journey of self-understanding.

We provide suggestions for AAC design based on our findings. The participants shared their experiences with going into shutdown, which typically limited their ability to use speech to communicate. Designing an AAC system targeting these times of shutdown could be beneficial for autistic adults. The participants also described wanting a way to indicate their communication ability to their communication partners. This would also indicate to others that the person using the AAC system is disabled, and the impact of this would also need to be explored.

We also share gaps in autism research that need further exploration. Several of the participants wrote about the impact of receiving an autism diagnosis later in life. This is an interesting and largely unexplored topic that needs further research. Similarly, the participants also shared their experiences with shutting down when overwhelmed or overstimulated. This created a unique set of communication needs that needs further exploration.

Finally, we show that online focus groups are a viable method for AAC and autism research. Making a focus group online, asynchronous, and text-based can make this methodology more accessible to autistic people and AAC users, but it is important to provide robust instruction on how to use the focus group platform. Our focus group allowed participants to interact with each other, discuss shared lived experiences, and gain a greater understanding of themselves through interacting with other autistic adults.


\bibliographystyle{ACM-Reference-Format}
\bibliography{references}

\appendix

\section{Weekly Prompts\label{appendix_prompts}}

\begin{itemize}
    \item Week 1:
    \begin{itemize}
        \item Please introduce yourself by sharing what forms of communication you use, where and when you use them, and anything else related to your communication methods and strategies you would like us to know.
        \item Please respond to at least one other participant!
    \end{itemize}
    \item Week 2:
    \begin{itemize}
        \item Could you share some stories about how you engage with your community?
        \item Could you give some examples of how your current communication methods help you engage with your community?
        \item Could you give some examples of how your current communication methods hinder you from engaging with your community?
    \end{itemize}
    \item Week 3:
    \begin{itemize}
        \item Do you use your current communication methods differently depending on who you’re communicating with? Your environment? The content of the conversation? If so, how?
        \item Does your communication style change depending on who you’re communicating with? Your environment? The content of the conversation? If so, how?
        \item How does your communication change based on the number of people you’re communicating with?
    \end{itemize}
    \item Week 4:
    \begin{itemize}
        \item What, if anything, do you wish that your community knew about communicating with autistic adults?
        \item What is the most important thing you think people in general should know about autistic adults communicating in social and community settings?
        \item What would be your ideal support system to help you communicate in community settings?
    \end{itemize}
    \item Week 5:
    \begin{itemize}
        \item What was your experience like participating in this online focus group?
        \item Were you able to effectively share your thoughts? Please explain what helped or hindered you.
        \item Were you able to engage with the other participants well? Please explain what helped or hindered you.
        \item If you could change one thing about how this focus group was run, what would it be and why?
        \item If you could keep one thing about how this focus group was run the same, what would it be and why?
    \end{itemize}
\end{itemize}

\end{document}